\documentclass[superscriptaddress,twocolumn,showpacs,amsmath,amssymb]{revtex4}

\usepackage{graphicx}

\renewcommand{\vec}[1]{\mbox{\boldmath $#1$}}

\begin{document}

\title{
Coexistence of BCS and BEC-like pair structures in halo 
nuclei}

\author{K. Hagino}
\affiliation{ 
Department of Physics, Tohoku University, Sendai, 980-8578,  Japan} 

\author{H. Sagawa}
\affiliation{
Center for Mathematical Sciences,  University of Aizu, 
Aizu-Wakamatsu, Fukushima 965-8560,  Japan}

\author{J. Carbonell}
\affiliation{Laboratoire de Physique Subatomique et de Cosmologie, 
F-38026 Grenoble Cedex, France} 

\author{P. Schuck}
\affiliation{ 
Institut de Physique Nucl\'eaire, CNRS, UMR8608, Orsay, F-91406, France} 
\affiliation{ 
Universit\'e Paris-Sud, Orsay, F-91505, France} 


\begin{abstract}
We investigate the spatial structure of the two-neutron wave function 
in the Borromean nucleus $^{11}$Li, using a three-body model
of $^9$Li $+n+n$, which includes many-body correlations stemming 
from the Pauli principle. 
The behavior of the neutron pair at different 
densities is simulated by calculating the two-neutron wave function at several 
distances between the core nucleus $^9$Li 
and the center of mass of the two
neutrons.  With this representation, 
a strong concentration of the neutron pair 
on the nuclear surface is for the first time quantitatively 
established
for neutron-rich nuclei. 
That is, the neutron pair wave function in 
$^{11}$Li has an oscillatory behavior 
at normal density, while it becomes a well localized single peak 
 in the dilute density region around the nuclear surface.
We point out that these features qualitatively 
correspond to 
the BCS and BEC-like structures 
of the pair wave function found in infinite nuclear matter.
\end{abstract}


\maketitle

Pairing correlations play a crucial role in many 
Fermion systems, such as liquid $^{3}$He, atomic nuclei, and
ultracold atomic gases \cite{BCS,FW71,BB05}. When the attractive
interaction between two Fermions is weak, the pairing correlation can 
be understood in terms of the well-known 
Bardeen-Cooper-Schrieffer (BCS) mechanism\cite{BCS}, showing a strong 
correlation in the momentum space. If the interaction is sufficiently
strong, on the other hand, one expects that two fermions form a 
bosonic bound state with condensation in the ground state of a many-body 
system 
\cite{E69,L80,NSR85,OG02,CSTL05}. The transition from the BCS-type pairing
correlations to the Bose-Einstein condensation (BEC) takes place
continuously as a function of the strength of the pairing interaction. 
This feature is often referred to as the BCS-BEC crossover. 

Recently, exploiting the Feshbach resonance with which the strength of 
effective interaction can be arbitrarily varied, the BCS-BEC
crossover has been experimentally realized for a gas of
ultracold alkali atoms \cite{GRJ03,J03,Z03}. 
This has stimulated a lot of subsequent works, not only in
condensed matter and atomic physics
\cite{CSTL05}, but also in nuclear and hadron physics \cite{M06,NA05} 
(see also Ref. \cite{BLS95}). 

Neutron-rich nuclei may 
manifest 
both BCS and BEC-like features. 
These nuclei are characterized by a dilute neutron density
around the nuclear surface, and one can investigate the
pairing correlation at several densities\cite{MMS05}, 
ranging from the normal density in the center of the nucleus to a dilute
density at the surface. 
In this connection, it is worth while to mention that 
Matsuo recently investigated the spatial structure 
of neutron Cooper pairs in low-density nuclear and neutron matters, and 
found the BCS-BEC crossover behavior in the pair wave function 
although the BEC limit is not reached because two neutrons 
are not bound in free space but only form a low-lying 
virtual state (see below) \cite{M06}. 
In  Ref. \cite{BLS95}, 
proton-neutron ($T$=0) Cooper pairs were also studied  in the same 
context. 
The strong density dependence of the nucleon-nucleon pseudo-potential, 
as well as the Pauli principle, 
are responsible for the crossover phenomenon. 

In this paper, we investigate 
the implication of the BCS-BEC crossover in 
{\it finite} neutron-rich nuclei. 
To this end, 
we particularly study the ground state wave function 
of a two-neutron halo nucleus, $^{11}$Li. 
This nucleus is known to be well described as 
   a three-body system 
consisting of two valence neutrons and the core nucleus $^{9}$Li
\cite{JJH90,BE91,EBH99,HS05,Barranco01,BP96}. 
Since both the $n$-$n$ and $n$-$^9$Li two-body subsystems are not bound, 
the $^{11}$Li nucleus 
is bound only as a three-body system. Such nuclei 
are referred to as Borromean nuclei, and have attracted 
a lot of attention \cite{Zhukov93,JRFG04}.
A strong di-neutron 
correlation as a consequence of the pairing interaction between the 
valence neutrons has been pointed out in $^{11}$Li \cite{BE91,HS05}, 
which has recently 
been confirmed experimentally in the low-lying dipole strength 
distribution\cite{N06}. 
This di-neutron correlation has a responsibility for the BEC-like 
behaviour in infinite nuclear matter, and thus, despite there is 
only one neutron pair, $^{11}$Li 
provides optimum circumstances
to investigate BCS and BEC-like 
features in finite nuclei. 

In order to study the pair wave function in 
$^{11}$Li,  we solve the following  three-body Hamiltonian 
\cite{EBH99,HS05}, 
\begin{equation}
H=\hat{h}_{nC}(1)+\hat{h}_{nC}(2)+V_{nn}+\frac{\vec{p}_1\cdot\vec{p}_2}{A_cm}, 
\label{3bh}
\end{equation}
where $m$ and $A_c$ are the nucleon mass and the mass number of the 
inert core nucleus, respectively. 
$\hat{h}_{nC}$ is the single-particle Hamiltonian for a valence 
neutron interacting with the core. 
We use a Woods-Saxon potential with a spin-orbit force 
 for the interaction in $\hat{h}_{nC}$.
The diagonal component of the recoil kinetic energy of the core
nucleus is included in 
$\hat{h}_{nC}$, whereas the off-diagonal part is taken into account
in the last term in the Hamiltonian (\ref{3bh}). 
The interaction between the valence neutrons $V_{nn}$ is 
taken as  a delta interaction whose strength depends
on the density of the core nucleus. 
This kind of pseudo-potential has been standard for nuclear pairing, 
see e.g., Refs. \cite{BE91,EBH99}. 
Assuming that the core density 
is described by a Fermi function, the pairing interaction reads 
\begin{equation}
V_{nn}(\vec{r}_1,\vec{r}_2)=\delta(\vec{r}_1-\vec{r}_2)
\left(v_0+\frac{v_\rho}{1+\exp[(R-R_\rho)/a_\rho]}\right), 
\label{vnn}
\end{equation}
where $R=|(\vec{r}_1+\vec{r}_2)/2|$. 
The density-dependent term is repulsive, and the strength of the 
interaction becomes {\it weaker} for {\it increasing} density. 
We use the same value for the parameters as in 
Refs. \cite{EBH99,HS05}, in which 
$R_\rho=2.935$ fm in the density-dependent term. 

The two-particle wave function $\Psi(\vec{r}_1,\vec{r}_2)$, 
where the coordinate of a valence neutron from the 
core nucleus is denoted  by $\vec{r}_i$, 
is obtained by
diagonalizing the  three-body Hamiltonian (\ref{3bh}) within a large
model space which is consistent with the $nn$ interaction,
$V_{nn}$. 
To this end, we expand the wave function 
$\Psi(\vec{r}_1,\vec{r}_2)$ 
with the  eigenfunctions of the single-particle
Hamiltonian $\hat{h}_{nC}$.  
In the expansion, we explicitly exclude those states which 
are occupied by the core nucleons, as in the original Cooper
problem\cite{BCS}. The ground state wave function 
is  obtained as the state with 
the total angular momentum $J=0$. 
We transform it to the
the relative and center of mass (c.o.m.) 
coordinates for the valence
neutrons, $\vec{r}=\vec{r}_1-\vec{r}_2$ and 
$\vec{R}=(\vec{r}_1+\vec{r}_2)/2$
 (see Fig. 1) \cite{CIMV84,TTD98,IAVF77}. 
To this end, we use the method of
Bayman and Kallio \cite{BK67}. 
That is, we first decompose 
the wave function into the 
total spin $S$=0 and $S$=1 components. 
The coordinate transformation is then 
performed  for the $S$=0 component, which is relevant to the
pairing correlation: 
\begin{equation}
\Psi^{S=0}(\vec{r}_1,\vec{r}_2)=
\sum_Lf_L(r,R)\,[Y_L(\hat{\vec{r}})Y_L(\hat{\vec{R}})]^{(00)}\,
|\chi_{S=0}\rangle,
\end{equation}
where $|\chi_{S=0}\rangle$ is the spin wave function. 

We apply this procedure to study the ground state wave function 
of the $^{11}$Li nucleus. 
We first discuss the probability of each $L$ component in the wave
function. Defining the probability as 
\begin{equation}
P_L\equiv\int_0^\infty r^2dr\,\int_0^\infty R^2dR\,|f_L(r,R)|^2,
\end{equation}
we find $P_L=0.578$ for $L=0$, ~0.020 for $L=2$, and 0.0045 for $L=4$. 
The $S$=0 component of 
wave function is thus largely dominated by the $L=0$ configuration 
\cite{JJH90}. The 
sum of the probabilities for $L=0,2,$ and 4 components 
is 0.6025, that is close to the 
$S=0$ probability in the total wave function, 0.606
\cite{EBH99,HS05}. 

\begin{figure}
\includegraphics[viewport=0 -180 150 300, scale=0.3,clip]{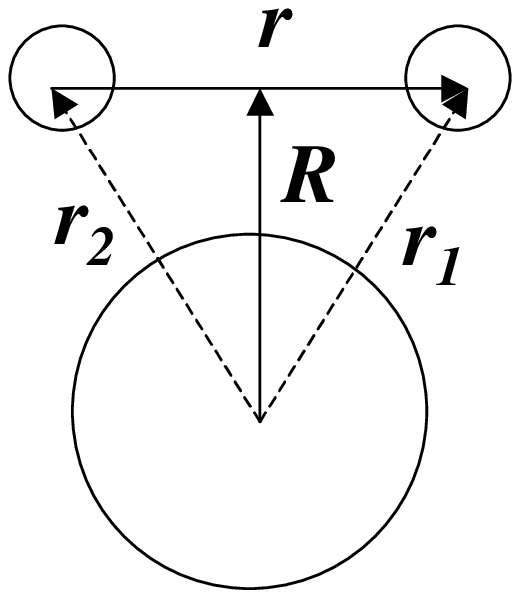}
\includegraphics[scale=1,clip]{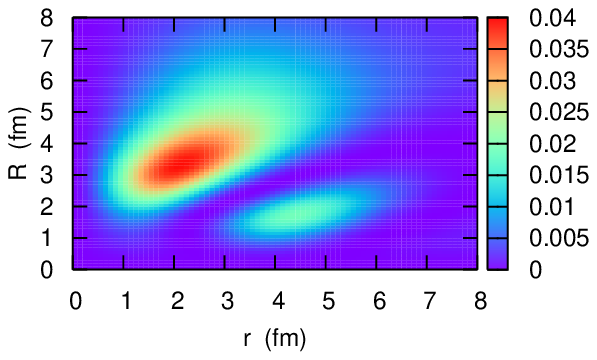}
\caption{(Color online) 
A two-dimensional (2D) plot for 
the  ground state 
two-particle wave function, $r^2R^2|f_{L=0}(r,R)|^2$, for
 $^{11}$Li. It is plotted as a function of the relative distance 
between two neutrons, $r$, and
the distance between the center of mass of the two neutrons 
and the core nucleus, $R$, as denoted 
in the inset. 
}
\end{figure}

Figure 1 shows the square of the two-particle wave function for the $L=0$ 
component. It is weighted with a factor of $r^2R^2$. 
One can clearly recognize the two peaked structure in the plot, one
peak at $(r,R)=(2.2,3.4)$ fm and the other at $(r,R)=(4.4,1.8)$ fm. 
These peaks correspond to the di-neutron and the cigar-like configurations 
discussed in Refs.\cite{BE91,HS05,Zhukov93}, respectively. 
Notice that the first peak is located at a small relative
distance between the neutrons and that the corresponding 
configuration is rather compact 
in the coordinate space.

\begin{figure}
\includegraphics[scale=0.35,clip]{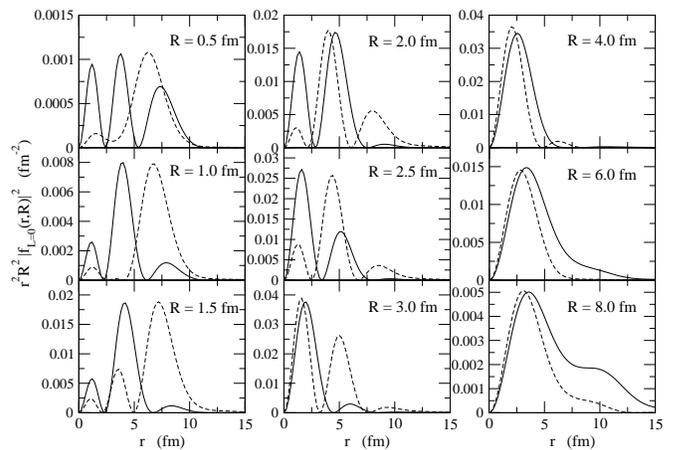}
\caption{
The  ground state 
two-particle wave functions, $r^2R^2|f_{L=0}(r,R)|^2$ as a function of 
the relative distance between 
the neutrons, $r$, at several distances $R$ from the core.
The solid lines correspond to the 
two-particle wave functions of $^{11}$Li, 
while the dashed lines denote   those of $^{16}$C. 
Notice the different scales on the 
ordinate in the various panels.}
\end{figure}

The $L=0$ wave functions of $^{11}$Li 
 for different values of $R$ are  
plotted in Fig.2 (solid line) as a function of $r$. 
The three-body wave function has not been presented in this way as 
far as we know, 
although the coordinate system ($\vec{r},\vec{R}$) has been 
employed in several previous calculations
\cite{JJH90,Zhukov93}. 
For comparison, those of 
$^{16}$C are also shown 
by the dashed lines with arbitrary scale. 
Since we consider the density-dependent contact interaction,
(\ref{vnn}), this is effectively equivalent to probing the wave
function at different densities. Let us first discuss the
wave function of $^{11}$Li. 
At $R=0.5$ fm, where the density 
is close to the normal density $\rho_0$, the two particle wave
function is spatially extended and oscillates inside the nuclear
interior. This oscillatory behavior is typical for a Cooper pair
wave function in the BCS approximation, and has in fact been found in 
nuclear and neutron matters at normal density $\rho_0$ (see Fig. 4 (f) in 
Ref. \cite{M06} as well as Fig. 4 in Ref. \cite{BLS95}). 
As in the infinite matter calculation\cite{M06}, 
the two-particle wave function has
a significant amplitude outside the first node at 2.4 fm. 
This is again a typical behavior of the BCS pair wave function. 
Notice that the core nucleus was assumed to be a point particle in 
Ref. \cite{JJH90}, and the oscillation of the pair wave function 
due to the Pauli principle is not seen there. 
As $R$ increases, the density $\rho$ decreases. 
The two-particle wave function then gradually deviates from the BCS-like 
behavior. 
At $R=3$ fm, the oscillatory behavior almost disappears and the 
wave function is largely concentrated inside the first node at 
$r\sim$ 4.5 fm. 
The wave function is compact in shape, indicating the strong 
di-neutron
correlation, typical for BEC 
when many such pairs are present. 
At $R$ larger than 3 fm, 
the squared wave function has 
essentially only one node, and the width of the peak gradually increases as a
function of $R$. This behavior is  qualitatively similar  
to  a local density approximation (LDA) picture of 
the pair wave function in the infinite 
matter \cite{M06}. 

The  present 
results also provide a unified picture of the di-neutron and the 
cigar-like configurations in Borromean nuclei. 
We have seen in Fig. 1 that, for $^{11}$Li, the former configuration 
corresponds to the 
peak around $r\sim$ 2.2 fm while the latter 
to the peak around $r\sim$ 4.4 fm. 
These correspond to the first and the second peaks of the solid lines 
in Fig. 2, respectively 
(see a typical case for $R=$2.0 fm). 
The transition from the BCS-like behavior of the wave
function to the BEC-like di-neutron  correlation 
shown in Fig. 2 thus suggests that the di-neutron and the cigar-like 
configurations are not independent of each other, but rather a 
manifestation of a single Cooper pair wave function probed at 
various densities. 

We have confirmed, using the same three-body model, 
that this scenario also holds for another 
Borromean nucleus $^6$He  
as well as for the non-Borromean neutron-rich 
nuclei $^{16}$C and  $^{24}$O. 
See the dashed line in Fig. 2 for $^{16}$C. 
The similarity with $^{11}$Li is striking. 
Namely, the oscillatory behavior is seen at small $R\le 3.0$ fm, while
a single compact peak appears at  $R\sim 4.0$ fm.
The surface condensation of the Cooper pair in several neutron-rich
nuclei has been discussed 
also in Refs. \cite{MMS05,Barranco01}, although these references 
use a coordinate system which does not remove the center of mass 
motion of two neutrons and the
surface condensation is less evident. 
We should mention that a similar, but less 
pronounced,  space correlation  
has already been mentioned earlier in Refs. 
\cite{CIMV84,TTD98}
for stable heavy nuclei. 
All of this indicates that the positioning of strongly coupled Cooper 
pairs with maximum probability in the nuclear surface is a quite 
common and general feature, that is enhanced significantly 
in the neutron-rich loosely-bound 
nuclei. 

\begin{figure}
\includegraphics[scale=0.4,clip]{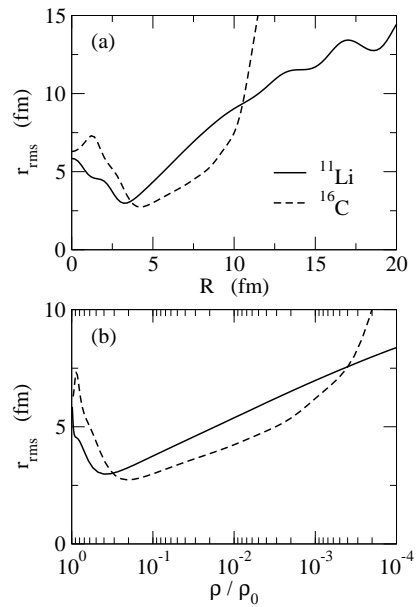}
\caption{
The root mean square distance $r_{\rm rms}$ for the neutron pair defined 
by Eq. (\ref{rms}). It is plotted as a function of the 
distance $R$ (Fig. 3(a)) and the density $\rho/\rho_0$ of the core nucleus
(Fig. 3(b)), where $\rho_0$ is the normal density of infinite nuclear 
matter. 
The solid and dashed lines are for $^{11}$Li and $^{16}$C,
respectively. 
}
\end{figure}

The transition from the BCS-type pairing to the BEC-type di-neutron 
correlation can also clearly be seen in the root mean square (rms)
distance of
the two neutrons. 
For a given value of $R$, we define the rms distance as 
\begin{equation}
r_{\rm rms}(R)\equiv 
\sqrt{\langle r^2_{nn}\rangle}\,(R)
=\sqrt{
\frac{\int_0^\infty r^4dr\,|f_0(r,R)|^2}{\int_0^\infty r^2dr\,|f_0(r,R)|^2}}. 
\label{rms}
\end{equation}
We plot this quantity in Fig. 3(a) as a function of $R$. 
In order to compare it with the rms distance in nuclear matter, we
relate the c.o.m. 
distance $R$ with the density $\rho$ using
the same functional form 
$\rho(R)/\rho_0
=[1+\exp((R-R_\rho)/a_\rho)]^{-1}$, 
as  used in the $nn$ interaction in Eq. (\ref{vnn}). 
Fig. 3(b) shows the rms distance thus obtained 
as a function of density $\rho$. 
The rms distance shows a distinct minimum at $\rho\sim 0.4\rho_0$
($R\sim$ 3.2 fm) in $^{11}$Li and $\rho\sim 0.2\rho_0$
($R\sim$ 4.2 fm) in $^{16}$C. 
This indicates that the strong di-neutron
correlations grow  in the two nuclei around these densities. 
Notice that the probability to find the two-neutron pair is maximal 
around this region (see Fig. 1). 
The behavior 
 of  rms distance 
as a function of density $\rho$ 
qualitatively well agrees with that in infinite matter 
(see Fig. 3 in Ref. \cite{M06}), although the absolute value of the
rms distance is much smaller in finite nuclei, 
since they are bound systems. 
A size shrinking effect has been found also for a
proton-neutron pair in infinite nuclear matter \cite{LS01} 
as well as in an old calculation for the 
$^{18}$O nucleus \cite{IAVF77}. 

Finally, let us discuss 
how the di-neutron wave function in $^{11}$Li is modified 
when approaching the $^{9}$Li core from infinite distance. 
It is known that a two-nucleon system in vacuum in the $^1S,~T=1$ 
channel ($L=S=0$)
has a virtual state around zero energy. 
In regularizing the rms distance using the method of Ref. \cite{VCL89},
it is obtained with the realistic Nijmegen potential \cite{Nijm} 
that  the virtual state has an extension of around 12 fm.
We therefore realize that in $^{11}$Li, in spite of being
a halo nucleus, 
the $nn$ singlet pair shows a dramatic change from its 
asymptotic behavior. 
Approaching the core nucleus $^9$Li, 
it shrinks down to an rms distance $r_{\rm rms}$ of only 2.6 fm at a
c.o.m. position of $R$=3.2 fm. 
At the same time, it has gained a maximum of binding. 
All this happens because of the well known Cooper pairing phenomenon. 
Pushing the $nn$ pair further to the center, it feels the 
increasing density of the neutrons of the core with which the $nn$ pair
needs to be orthogonal. 
Therefore, approaching the center, the Cooper pair again looses binding 
and thus increases in size. 
What is surprising is that there exists such a well pronounced radius 
in the surface where the Cooper pair has minimum extension
and highest probability of presence (see Figs. 1-3). 
This seems a quite general feature common to many nuclei with
well developed pairing correlations as shown in Fig. 2 (see also 
Ref. \cite{PSS07}). 

In summary, we studied the two-neutron wave function 
in the Borromean nucleus $^{11}$Li by using a three-body model with 
a density-dependent pairing force, 
and discussed its relation to the Cooper pair wave function
in infinite matter. We explored the spatial distribution of 
the two neutron wave
function as a function of 
the center of mass distance $R$ from the core nucleus, that allows 
an optimal representation of the physics. 
We found that the  structure
 of the two-neutron wave function alters drastically as $R$
 is varied, in a qualitatively similar way to that for the infinite matter. 
We  also showed that the relative distance  
between  the two neutrons  
scales consistently to that in the infinite 
matter as a function of density. 
These features show 
the same characteristics of coexistence of BCS and BEC-like behaviors 
found in infinite nuclear and neutron
matters. 

\medskip

We thank M. Matsuo, N. Sandulescu, and K. Kato 
for useful discussions and 
informations. 
K.H. thanks the theory group of IPN Orsay, where this work was
initiated, for its warm hospitality and financial support. 
This work was supported by the Japanese
Ministry of Education, Culture, Sports, Science and Technology
by Grant-in-Aid for Scientific Research under
the program numbers (C(2)) 16540259 and 16740139.

\end{document}